\documentclass[prl,superscriptaddress,showpacs,twocolumn,longbibliography]{revtex4-1}

\usepackage{hyperref}

\usepackage{color}
\usepackage{amsmath,amsthm,amssymb}
\usepackage{graphicx}
\usepackage{bm}
\usepackage{subfig}

\newcommand{\er}[1]{Eq.~\eqref{#1}}

\newcommand{\era}[2]{Eqs.~(\ref{#1}) and (\ref{#2})}

\newcommand{\bra}[1]{\langle #1 |}
\newcommand{\ket}[1]{| #1 \rangle }

\newcommand{\Egs}{E_{\rm GS}(s)}
\newcommand{\brags}{\langle {\Phi_{s0}} |}
\newcommand{\ketgs}{| {\Phi_{s0}} \rangle}

\newcommand{\SE}{S_{\rm E}}
\newcommand{\Tr}{{\rm Tr}}

\begin{document}

\title{Using matrix product states to study the dynamical large deviations
of kinetically constrained models} 

\author{Mari Carmen Ba\~nuls}
\affiliation{Max-Planck-Institut f\"ur Quantenoptik, Hans-Kopfermann-Str.\ 1, D-85748 Garching, Germany}
\affiliation{Munich Center for Quantum Science and Technology (MCQST), Schellingstr. 4, D-80799 M\"unchen}
\author{Juan P. Garrahan}
\affiliation{School of Physics and Astronomy, 
University of Nottingham, Nottingham, NG7 2RD, UK}
\affiliation{Centre for the Mathematics and Theoretical Physics of Quantum Non-Equilibrium Systems,
University of Nottingham, Nottingham, NG7 2RD, UK}

\date{\today}

\begin{abstract}
Here we demonstrate that tensor network techniques --- originally devised for the analysis of quantum many-body problems --- are well suited for the detailed study of rare event statistics in kinetically constrained models (KCMs). As concrete examples we consider the Fredrickson-Andersen and East models, two paradigmatic KCMs relevant to the modelling of glasses. We show how variational matrix product states allow to numerically approximate --- systematically and with high accuracy --- the leading eigenstates 
of the tilted dynamical generators which encode the large deviation statistics of the dynamics. Via this approach we can study system sizes beyond what is possible with other methods, allowing us to characterise in detail the finite size scaling of the trajectory-space phase transition of these models, the behaviour of spectral gaps, and the spatial structure and ``entanglement'' properties of dynamical phases. We discuss the broader implications of our results.
\end{abstract}

\maketitle 

\noindent
{\bf \em Introduction.--} Dynamics equipped with local kinetic constraints provides a general mechanism for slow cooperative relaxation  \cite{Palmer1984,Fredrickson1984,Jackle1991,Kob1993}. Kinetically constrained models (KCMs) --- of which the Fredrickson-Andersen (FA) \cite{Fredrickson1984} and East \cite{Jackle1991} facilitated spin models are the simplest exponents --- give many insights into the nature of glass forming systems, in particular by showing that systems with simple thermodynamics can have rich, spatially fluctuating and slow dynamics \cite{Garrahan2002}. (For reviews on the glass transition see \cite{Binder2011,Berthier2011,Biroli2013}, and on KCMs see \cite{Ritort2003,Garrahan2011,Garrahan2018}.) Beyond glasses, classical KCMs (and related deterministic models \cite{Prosen2016,Inoue2018,Prosen2017,Klobas2018,Buca2019}) are  relevant to the problem of operator spreading in quantum systems \cite{Nahum2017,Rowlands2018,Chen2018,Gopalakrishnan2018,Knap2018,Tran2018,Gopalakrishnan2018b,Alba2019} and to non-equilibrium dynamics of ensembles of Rydberg atoms \cite{Lesanovsky2013,Urvoy2015,Valado2016}, while quantum KCMs provide a template for  complex non-equilibrium dynamics under unitary evolution in the absence of disorder \cite{Horssen2015,Smith2017,Lan2018,Turner2018}. 

To characterise dynamics it is natural to study ensembles of stochastic trajectories, just like one does in equilibrium statistical mechanics with ensembles of configurations. For long-times one can then apply the methods of dynamical large deviations (LDs) \cite{Touchette2009} to compute quantities that play the role of thermodynamic potentials for the dynamics. For the case of KCMs this ``thermodynamics of trajectories'' approach reveals the existence of a first-order phase transition in the space of trajectories between {\em active} and {\em inactive} dynamical phases, indicative of the singular change when fluctuating away from typical behaviour \cite{Garrahan2007,Garrahan2009}. Many other systems have been also shown to have similar LD transitions, see e.g.\ \cite{Lecomte2007,Appert-Rolland2008,Hedges2009,Speck2012,Weber2013,Espigares2013,Jack2015,Karevski2017,Baek2017,Oakes2018}. Understanding the phase structure of the dynamics is clearly as important in dynamical problems as it is in static ones.

The standard way of accessing LD statistics of a dynamical observable is by computing its scaled cumulant generating function (SCGF) --- see below for definitions --- from the largest eigenvalue of an appropriate deformation, or {\em tilting}, of the generator of the dynamics \cite{Touchette2009,Garrahan2018}. Except for the handful of non-trivial cases in which it can be calculated exactly \cite{Appert-Rolland2008,Buca2019}, obtaining the SCGF by diagonalising the tilted generator is 
only possible for small system sizes. To access the LD behaviour for larger sizes one has to resort to numerical methods for sampling rare trajectories based on splitting/cloning, importance sampling or optimal control \cite{Giardina2006,Cerou2007,Lecomte2007b,Nemoto2016,Hedges2009,Ray2018,Klymko2018,Ferre2018}. 

By exploiting the similarity between tilted generators and quantum Hamiltonians, here we show how to use variational matrix product states (MPS) to compute numerically with high accuracy (and precise control on errors) leading eigenvalues and eigenstates of the tilted generator for system sizes way beyond those accessible through other methods. We study in detail the FA and East models, focusing on the finite size scaling of their active-inactive phase transitions and the spatial structure that emerges in the dynamical phases. 
While in certain special cases MPS can be used to obtain exact LD statistics, such as in simple exclusion processes \cite{Derrida1998,Gier2011,Lazarescu2011,Gorissen2012,Crampe2016}, hard core brownian particles \cite{Lapolla2018}, and certain cellular automata \cite{Buca2019}, the systematic application of numerical MPS methods to stochastic lattice systems has been limited \cite{Gorissen2009}. Our results for KCMs --- together with the very recent ones \cite{Helms2019} for simple exclusion processes --- show the potential of numerical tensor network methods for the detailed study of dynamical fluctuations in stochastic dynamics.

\begin{figure*}[t]
\centering
\includegraphics[width=\textwidth]{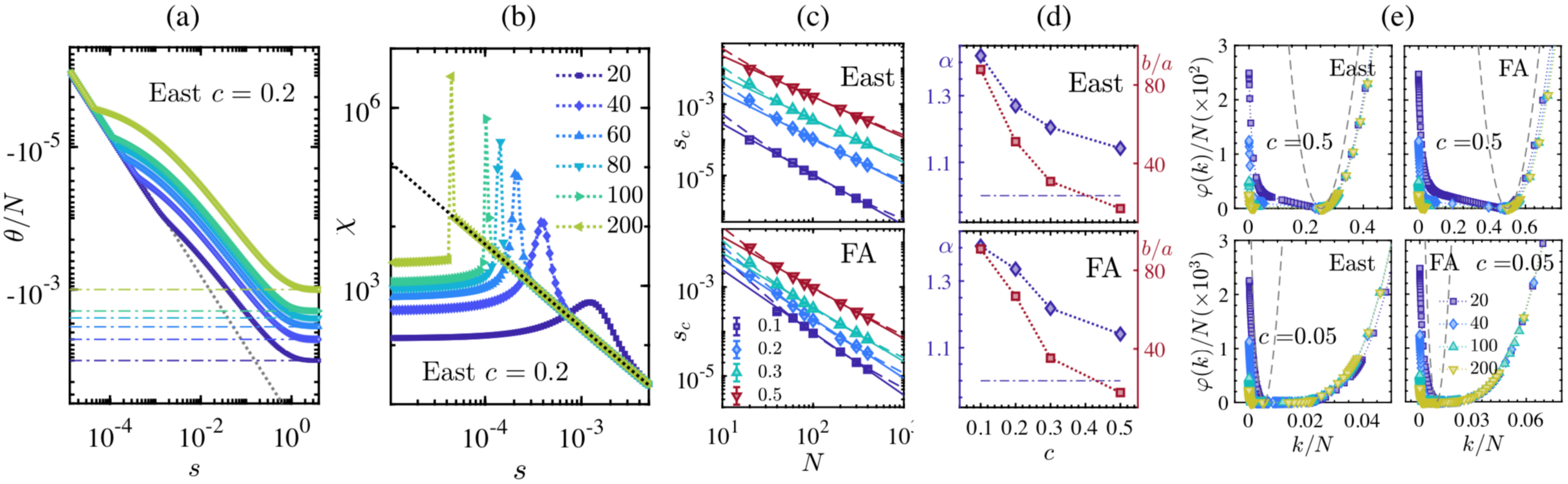}
\caption{{\bf Finite size scaling of trajectory transition.} (a) SCGF $\theta(s)/N$ as a function of $s$ for the East model ($c=0.2$) for system sizes $N=20$ to $N=200$. 
The critical $s_c(N)$ can be obtained from the (extrapolated) crossing of the first two energy levels.
The dot-dashed lines correspond to the asymptotic values $\theta(s\to\infty)=-c$. 
(b) The corresponding dynamical susceptibilities, $\chi(s) = \theta''(s)$,
exhibit a peak at $s_c(N)$ that gets narrower and higher as $N$ increases.
For $s>s_c(N)$  we find an almost universal behavior $\chi \propto s^{-\gamma}$ with $\gamma \approx 1.4$.
(c) $s_c(N)$ as a function of $N$ for $N \in [20,400]$ and various equilibrium concentrations $c$ in the East model (top) and FA model (bottom). 
As expected the data is compatible with $\lim_{N \to \infty} s_c(N) \to 0$, but $s_c$ appears to scale as $s_c(N) \propto N^{-\alpha}$ with $\alpha > 1$ (full lines are power-law fits; for comparison we also show fits to $a/N+b/N^2$, dashed). 
(d) The scaling exponents $\alpha$ (blue diamonds) and fitting parameters $b/a$ (red squares) as a function of $c$ (top, East model; bottom, FA model). The departure from $1/N$ scaling (dotted-dashed) appears to be more pronounced the lower the $c$ is. 
(e) Rate functions $\varphi(k)$ for $N \in [20,200]$ for the East model (left) and FA model (right) at $c=0.5$ (top) and $c=0.05$ (bottom). The dashed lines correspond to Poisson distributions with average $\langle k \rangle = -\theta'(0)/N$.  
}
\end{figure*}

\smallskip 

\noindent {\bf \em  FA and East models.--} The FA \cite{Fredrickson1984} and East \cite{Jackle1991} models are defined in terms of binary variables, $\{ n_i =0,1 \}_{i=1}^N$, on the sites of a one dimensional lattice of size $N$, with single-spin flip dynamics subject to a kinetic constraint such that a spin can flip up (with rate $c$) or down (with rate $1-c$) {\em only} if either nearest neighbour is in the up state (FA model) or {\em only} if the leftmost nearest neighbour is in the up state (East model). The generators for the corresponding continuous time Markov chains are \cite{Ritort2003,Garrahan2011,Garrahan2018}
\begin{align}
W^{\rm FA} &= \sum_i
\left( n_{i-1} + n_{i+1} \right)
\left[ c \sigma_i^+ + (1-c) \sigma_i^-
\right.
\nonumber \\
& \left.
 - c (1-n_i) - (1-c) n_i \right] ,
\label{WFA}  \\
W^{\rm East} &= \sum_i
n_{i-1}
\left[ c \sigma_i^+ + (1-c) \sigma_i^- \right.
\nonumber \\
& \left.
- c (1-n_i) - (1-c) n_i \right] ,
\label{WEast}
\end{align}
where $\sigma_i^\pm$ flips the site $i$ up/down, and the factor in front of the square brackets is the kinetic constraint. In this formulation the master equation is $\partial_t | P \rangle = W | P \rangle$, where $| P \rangle$ is the probability vector over configurations. 

We consider {\em open boundary conditions} which formally corresponds to setting $n_0 = n_{N+1}=0$ in \era{WFA}{WEast}. This is the best setup for the MPS method we use below.  Due to the kinetic constraints configuration space can be disconnected, and we consider the dynamics within the largest ergodic component: the set of all configurations with at least one up site for the FA model, and all the configurations with fixed $n_1=1$ for the East model. 

The dynamics has as stationary distribution $| P_{\rm eq} \rangle$ given by a projection of the product state $| c \rangle^{\otimes N}$, where $|c\rangle = (1-c) |0\rangle + c |1\rangle$, into the relevant ergodic component, giving
\begin{align}
| P_{\rm eq}^{\rm FA} \rangle &= [| c \rangle^{\otimes N} - (1-c)^N | 0 \rangle^{\otimes N}]/[1 - (1-c)^N] ,
\label{PeqFA} \\
| P_{\rm eq}^{\rm East} \rangle &= |1 \rangle \otimes | c \rangle^{\otimes N-1} .
\label{PeqEast}
\end{align}
These are the equilibrium distributions with energy $E = \sum_i n_i$ at inverse temperature $\ln (1-c)/c$ in the corresponding ergodic components.

\begin{figure*}[t]
\centering
\includegraphics[width=\textwidth]{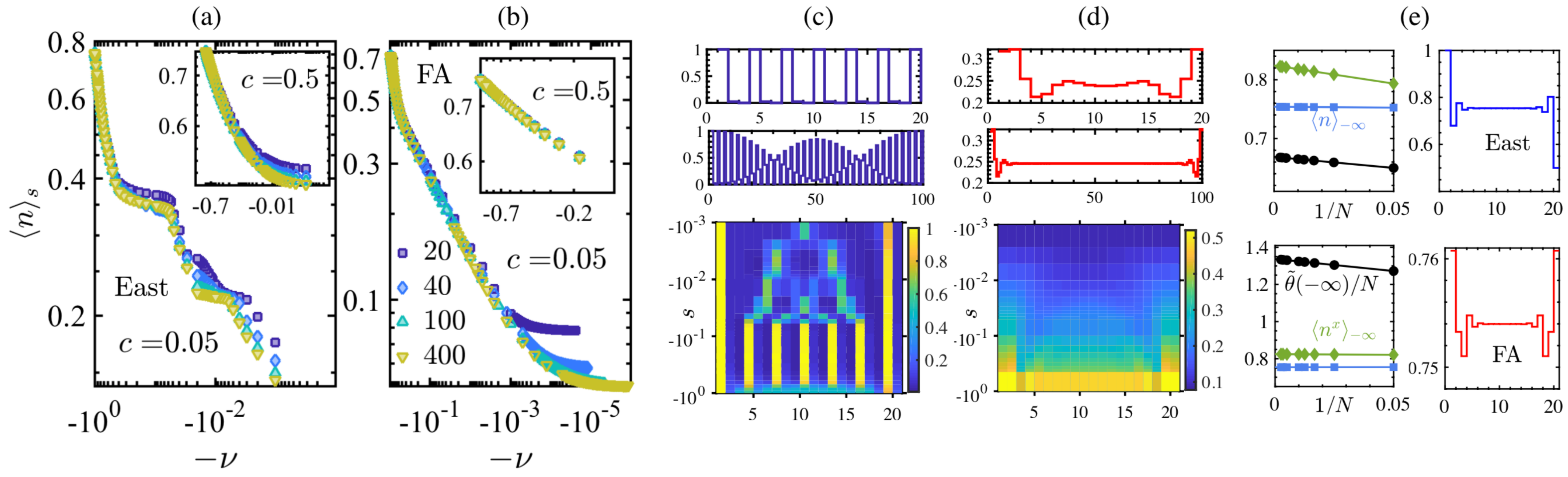}
\caption{{\bf Structure of active phase.} (a) Mean density $\langle n \rangle_s$ in the active phase, $s<0$,  in the East model for the case $c=0.05$ (shown as function of $-\nu = e^s-1$). For small $c$ the plateau structure of the density is evident (as compared to $c=0.5$ in the inset). (b) Same for the FA model, where the plateaus are absent.
(c) Density profile of the ground state of $H_s$ at $\nu = 0.081$ ($s=-0.0845$) for the East model at $c=0.05$ for sizes $N=20,100$ (top and middle panels) and 
density profiles across the active phase for $N=20$ (bottom panel). (d) Same for the FA model at $c=0.05$. 
For the East model the state has pronounced anticorrelations which are absent in the FA case model. 
(e) Extreme limit of the active phase, $s \to -\infty$, for the East and FA models (top and bottom, respectively). In the panels on the right the symbols show the rescaled 
$\tilde{\theta}(s=-\infty)/N:=e^s\theta(s=-\infty)/[N\sqrt{c(1-c)}]$ (black circles), $\langle n \rangle_{s=-\infty}$ (blue squares) and 
$\langle n^x \rangle_{s=-\infty}$ (green diamonds) for $N \in [20,400]$. The lines are fits to $a/N + b$ to extract the values in the thermodynamic limit: $\lim_{N \to \infty} \theta(s=-\infty)/N,\langle n^x\rangle_{s=-\infty},\langle n \rangle_{s=-\infty} = 0.67,0.82,0.75$ (East) and $1.34,0.82,0.75$ (FA). The right panels show (for $N=20$) that the density profiles at $s=-\infty$ are uniform, up to boundaries, in both models.
}
\end{figure*}

\smallskip

\noindent
{\bf \em Dynamical LDs and tilted generators.--} As trajectory observable we will consider the {\em dynamical activity} \cite{Lecomte2007,Garrahan2007,Baiesi2009b}, given by the total number of configuration changes $K(\omega_t)$ (i.e., number of spin flips) in a trajectory $\omega_t$ of time extent $t$. For large $t$ the distribution of $K$ obeys a LD principle, $P_t(K) = \langle \delta[ K(\omega_t) - K ] \rangle \approx e^{-t \varphi(K/t)}$, where $\varphi(x)$ is the LD rate function \cite{Touchette2009}. 
The corresponding moment generating function $Z_T(s) = \langle e^{-s K(\omega_t)} \rangle$ also obeys a LD principle, $Z_T(s) \approx e^{t \theta(s)}$, where $\theta(s)$ is the {\em scaled cumulant generating function} (SCGF), whose
 derivatives at $s=0$ give the cumulants of $K$ (scaled by $t$) \cite{Touchette2009}. The LD functions are connected by a Legendre transform, $\theta(s) = - \min_k \left[ s k + \varphi(k) \right]$ \cite{Touchette2009} and play the role of thermodynamic potentials for trajectories. 

The SCGF can be obtained from the largest eigenvalue of a tilted generator, 
$W_s$ 
\cite{Touchette2009}. For the case of the dynamical activity, the tilt corresponds to multiplying the off-diagonal terms of $W$ by a factor $e^{-s}$ \cite{Garrahan2007,Lecomte2007}. Since the dynamics obeys detailed balance, the generators can be made hermitian by a similarity transformation which is independent of $s$ \cite{Garrahan2009}. That is, if we define $H_s = - Q^{-1} W_s Q$, where $Q$ is a diagonal matrix with elements $\bra{{\bf n}} Q \ket{{\bf n}} = (1-c)^{N/2} [c/(1-c)]^{\sum_i n_i/2}$ 
in the configuration basis $\{ | {\bf n} \rangle \}$, we get
\begin{align}
H^{\rm FA}_s & = - \sum_i
\left( n_{i-1} + n_{i+1} \right)
\label{HFA} \\
& 
\times 
\left[ e^{-s} \sqrt{c(1-c)} \sigma_i^x - c (1-n_i) - (1-c) n_i \right] ,
\nonumber \\
H^{\rm East}_s &= - \sum_i
n_{i-1}
\left[ e^{-s} \sqrt{c(1-c)} \sigma_i^x 
\right.
\label{HEast} \\
&
\left.
\phantom{\sqrt{c(1-c)}}
- c (1-n_i) - (1-c) n_i \right] ,
\nonumber
\end{align}
The SCGF therefore corresponds to (minus) the ground state energy of $H_s$, 
\begin{equation}
\theta(s) = - \Egs .
\label{Egs} 
\end{equation}

The relation between the ground state $\ketgs$ of the tilted Hamiltonian, $H_s \ketgs = \Egs \ketgs$, and the left $\langle L_s |$ and right $| R_s \rangle$ leading eigenvectors of the tilted generator, $W_s | R_s \rangle = \theta(s) | R_s \rangle$, $\langle L_s | W_s = \langle L_s | \theta(s)$, is
\begin{equation}
\ketgs = \sum_{\bf n} \sqrt{l_{\bf n}(s) r_{\bf n}(s)} \, | {\bf n} \rangle
\label{eq:gs-ev}
\end{equation}
where $l_{\bf n}(s) = \langle L_s |{\bf n} \rangle$ and $r_{\bf n}(s) = \langle {\bf n} | R_s \rangle$. The aim now is to compute $\Egs$ and $\ketgs$ for \era{HFA}{HEast}.

\smallskip

\noindent
{\bf \em Variational MPS method.--} For a lattice of $N$ $d$-dimensional quantum systems, 
a MPS \cite{Perez-Garcia2007} is a vector 
$|\Psi\rangle=\sum_{i_1,\ldots i_N=1}^d \mathrm{tr}\left ( A_1^{i_1} A_2^{i_2}\ldots A_N^{i_N} \right ) |i_1 i_2 \ldots i_N\rangle$, where $i_k$ labels a local basis of the $k-$th subsystem, and
each $A_k$ is a rank-$3$ tensor of dimensions $d\times D\times D$~\footnote{In the case of open boundary conditions,
as used in this work, the first and last tensors reduce to rank-$2$ tensors of dimensions $d\times D$.}. Such a state is described by $O(dND^2)$ parameters. The {\em bond dimension} $D$ limits the entanglement 
of the state. More precisely, in an MPS of bond dimension $D$, for any subchain $A$ the {\em entanglement entropy} (defined as $\SE = - \Tr_A \rho_A \log \rho_A$, where $\rho_A = \Tr_{N \setminus A} |\Psi\rangle \langle \Psi|$ \cite{Nielsen2011}) is upper-bounded by $\SE\leq 2 \log D$, independent of the subchain length.
Namely, MPS satisfy an entanglement \emph{area law} \cite{Eisert2010}, and conform a hierarchy of increasingly entangled states, with $D=d^{N/2}$ sufficing to describe the whole Hilbert space.

Conversely, MPS can efficiently approximate states that satisfy an area law~\footnote{Strictly speaking, the statement holds for states which fulfill an area law in Renyi entropies $S_{\alpha}=\log(\mathrm{tr} \rho^{\alpha})/(1-\alpha)$ with $0<\alpha<1$ \cite{Schuch2008}.}, such as ground states of gapped local Hamiltonians. They thus are the basis for numerical methods like the celebrated
density matrix renormalization group (DMRG) 
algorithm~\cite{White1992} which can be understood as a variational minimization of energy over MPS~\cite{Vidal2003,Verstraete2004,McCulloch2007,Verstraete2008,Schollwoeck2011},
by sequientially optimizing a single tensor, while keeping the rest constant, and
iteratively sweeping over the chain until convergence
\footnote{Notice that it is also possible to define MPS directly in the thermodynamic limit, and optimize them numerically 
with appropriate methods~ \cite{Verstraete2008,Schollwoeck2011}.}.
Formulated in terms of tensor networks this algorithm allows a number of extensions,
including simulating dynamics, and the calculation of a few excited states above the ground state.

We apply this strategy to find MPS approximations to the ground state and first excitations of the Hamiltonians \eqref{HFA} and \eqref{HEast}. 
In this case, $d=2$ and the basis is $\{ | {\bf n} \rangle \}$.
As we show below, MPS with $D \ll 2^N$ provide accurate approximations for systems sizes at  an order of magnitude larger than those accessible by other methods 
\footnote{
For details on the MPS numerics, their convergence, and for the comprehensive set of results for both the FA and East models, see Supplemental Material.}.

\begin{figure}[t]
\centering
\includegraphics[width=\columnwidth]{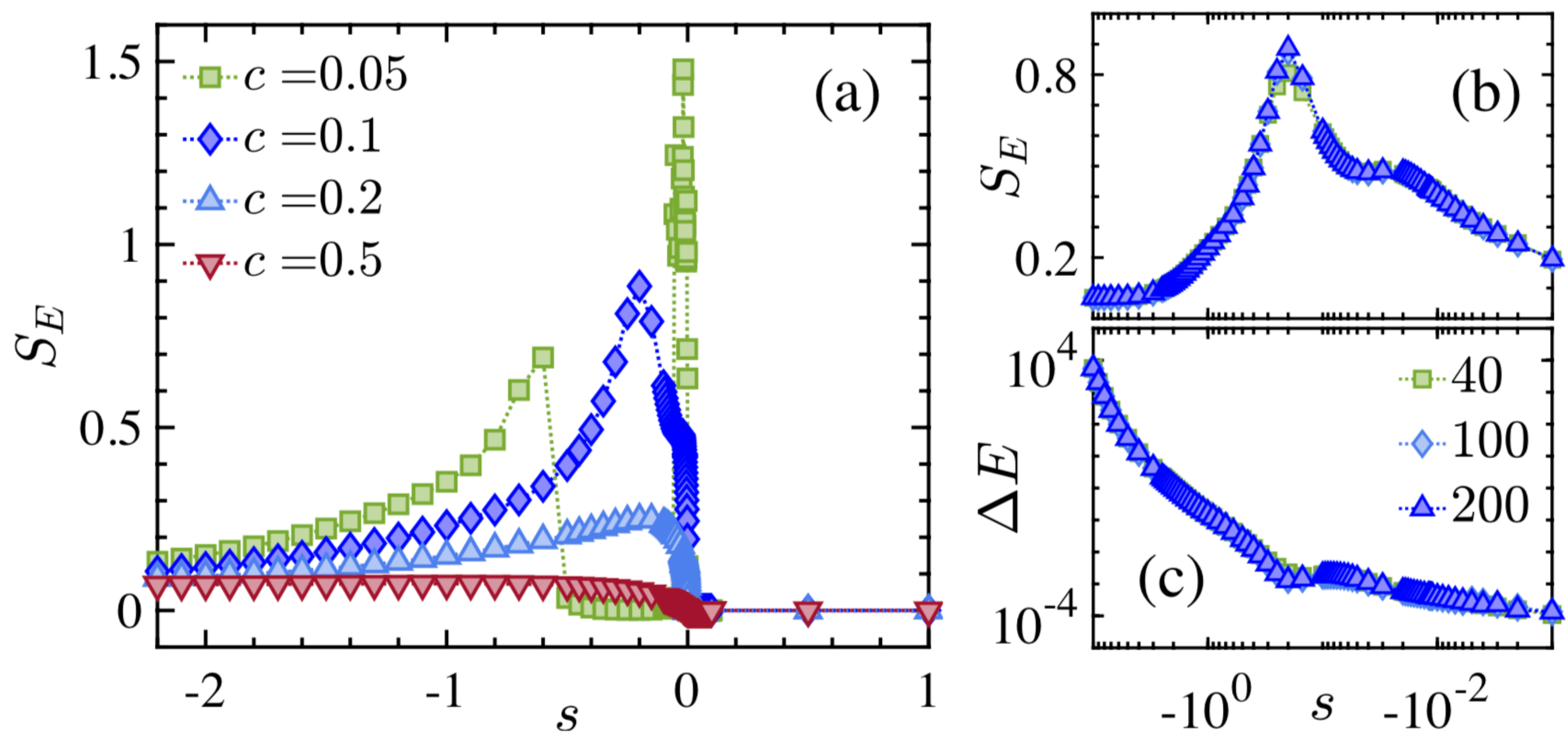}
\caption{{\bf Entanglement.} (a) Half-chain entanglement entropy $S_E$ of the ground state of $H_s$ as a function of $s$ for $c=0.5,0.1,0.05$ in the East model at $N=200$. (b) $S_E$ for $s<0$ for $c=0.1$ at various sizes $N$. The peak is correlated with the change in shape of the spectral gap $\Delta E$ of $H_s$ shown in (c).}
\end{figure}

\smallskip

\noindent
{\bf \em Results. Finite size scaling of active-inactive trajectory transition.--} 
The key property of KCMs like the FA and East is their first-order phase transition between an {\em active} phase for $s<0$ and {\em inactive} dynamical phase at $s>0$ \cite{Garrahan2007,Garrahan2009}, manifested in a first-order singularity in the SCGF in the limit of $N \to \infty$. Like for all phase transitions, to characterise the transition and its associated fluctuations, it is necessary to understand how the singularity is approached as the system size increases. Theoretical and numerical considerations 
 \cite{Bodineau2012,Bodineau2012b,Nemoto2017} suggest that
for finite $N$ the (rounded) transition occurs at $s_c(N) > 0$ (i.e.\ typical dynamics, $s=0$, is perturbatively connected to the active phase), and $s_c(N) \to 0^+$ as $1/N$. These predictions can be tested with our MPS method.

Figure 1(a) shows (minus) the energy density $-\Egs/N = \theta(s)/N$ of the MPS solution as a function of $s$. The transition at $s_c(N)$ occurs where the two branches cross. 
The leftmost branch is linear in $s$ and proportional to $N$, corresponding to the linear response for $s \gtrsim 0$ (grey dashed line). 
The rightmost branch is nonlinear, connecting the regime at $s \gtrsim 0$ to the asymptotic $\theta(\infty) = -c$. The corresponding susceptibility $\chi_s = \theta''(s)/N$ shows a diverging peak at $s_c(N)$, see Fig.~1(b) \cite{Note4}.

We can estimate the location of $s_c(N)$ from the susceptibility peak. For both models we find a departure from the expected $1/N$ scaling. Figure 1(c) shows that $s_c(N)$ can be fit to a power law, $s_c(N) \propto N^{-\alpha}$ with $\alpha > 1$ throughout. An alternative is that this discrepancy is due to subleading corrections to $1/N$, see Fig.~1(c) (dashed lines), and Fig.~1(d) for the dependence of the scaling parameters with $c$. We also show in Fig.~1(e) the broadening with $N$ of the LD rate function, indicative of the first order transition \cite{Garrahan2007,Garrahan2009}. For more details on the finite size scaling analysis including comparison with the predictions of the Ref.~\cite{Bodineau2012} see \cite{SM}.

\smallskip

\noindent
{\bf \em Structure of active phase.--} While both models have similar active-inactive  transitions, their active phases differ. 
Figures 2(a,b) show the average density of excitations, $\langle n \rangle_s = N^{-1} \sum_{i=1}^N \brags n_i \ketgs$, 
in the MPS that approximates the ground state of $H_s$ for $s<0$. In the East model and for small $c$,  $\langle n \rangle_s$ shows a series of plateaus as $s$ becomes more negative,
as predicted in Ref.~\cite{Jack2013}.
These plateaus are absent in the FA model at the same $c$, Fig.~2(b),
and also when the equilibrium concentration $c$ is high, see insets to Figs.~2(a,b).

Figures 2(c,d) show the difference in spatial structure of the active phases. The top two panels in Fig.~2(c,d) give the density profile at $s=-0.0845$ ($\nu = 0.081$) corresponding to the plateau in Fig.~2(a) with density $\langle n \rangle_s \approx 1/3$. For the East model, Fig.~2(c, top two panels), the state is anticorrelated in space, with an occupied site followed by two nearly empty ones. This is evident in the $N=20$ case, shown in the figure, while for $N=100$ we also observe a longer ranged modulation of this pattern \cite{Note4}. In contrast, in the FA model the density is essentially uniform, Fig.~2(d, top two panels). This difference in structure is present throughout the $s<0$ phase, see bottom panels of Fig.~2(c,d). 

We can also characterise the extreme active limit $s \to -\infty$.
We find that a MPS of $D \sim O(10)$ is enough to obtain a very precise approximation to the  ground state over the whole range of sizes computed, $N\in[20,\, 400]$. We can then extrapolate to $N \to \infty$. We obtain, Fig.~2(e), for the limiting SCGFs
 $\lim_{N \to \infty} \lim_{s \to -\infty} e^{s} \theta_{\mathrm{E}}(s)/[ N \sqrt{c(1-c)}]\approx 0.6687$ for the East and $1.337$ for the FA model,
 while the densities are the same in both models,
 namely
 $\lim_{N \to \infty} \langle n \rangle_{-\infty} \approx 0.754$ and 
 $\lim_{N \to \infty} \langle n^x \rangle_{-\infty} \approx 0.824$ (where $n^x$ is up to constants the ``transverse'' magnetisation, $2 n^x = 1 - N^{-1} \sum_{i=1}^N \sigma_i^x$). The panels on the right of Fig.~2(e) show that the corresponding density profiles are essentially flat in this limit
\footnote{
The GS in the limit $s\to -\infty$ seems to be gapped and with low entanglement ($D\sim 10$ provides a very good approximation \cite{Note4}). The GS energy of the FA in this limit is almost exactly twice the one for East, and the overlap of their states is very high, suggesting they have similar GS, or rather the FA one is the superposition of that of the East and the reflected ``West'' model.}.

\smallskip

\noindent
{\bf \em Entanglement.--} The states at $s \neq 0$ have spatial correlations absent in equilibrium ($s=0$) and which varies with $s$. This can be quantified via their entanglement entropy, which together with other quantum information measures can capture changes in dynamical behaviour that might escape classical order parameters \cite{Castelnovo2010}. The entanglement entropy is easily computed for a state in MPS form.
Figure 3(a) shows the half-chain $S_E$ of the state $\ketgs$
as a function of $s$ in the East model at size $N=200$. It is zero in the equilibrium state, cf.~\er{PeqEast}, and very small in the inactive phase, where the leading eigenvector is close to a product state of all sites empty in the bulk. For $s<0$ it shows interesting structure, as expected from the spatial correlations of Fig.~2. 
In Fig.~3(b) we notice that the maximum of $\SE$ does not seem to scale with system size. Thus, in the language of quantum many-body systems, the ground state fulfils an area law. This is also the case for other entropic quantities \cite{Note4}, which justifies the accuracy of the MPS approximation. 

The peak in $S_E$ nevertheless is sensitive to changes in the structure of the active phase. Fig.~3(c) shows the corresponding gap between $\Egs$ and the eigenvalue of the first excited state: its $s$ dependence changes at a value of $s$ located by the peak in $S_E$. (Note also that the gap is has no significant $N$ dependence.)
The maximum of the entropy depends on the value of $c$, and we find a larger peak for smaller values, corresponding to richer structure in the 
active phase, see Fig.~3(a) and \cite{Note4}. 

Even if the entanglement is low throughout the phase diagram, cf.\ Fig.~3(a), this does not guarantee that the variational method will easily find an MPS approximation. In fact, we find that both for the region close to the phase transition at $s=0$ and for the values of $s$ where $\SE$ shows a peak, cf.\ Fig.~3(a,b), the numerical convergence is slower than would have been expected. We believe this is a consequence of how the spectrum of the Hamiltonian changes when approaching these regimes \cite{Note4}.

\smallskip

\noindent
{\bf \em Conclusions.--} As we have shown here, 
the MPS methods often employed in quantum many-body problems \cite{Schollwoeck2011}, are also well suited for the study of the dynamical generators of classical stochastic systems
\cite{Derrida1998,Gier2011,Lazarescu2011,Gorissen2012,Crampe2016,Lapolla2018,Buca2019,Gorissen2009,Prosen2016,Inoue2018,Prosen2017,Klobas2018,Helms2019}. We focused on the LD statistics of KCMs such as the FA and East models, and showed how variational MPS approximations allow to efficiently access system sizes which are larger by an order of magnitude compared to previous studies, thus providing detailed information about the properties of the transitions in these models and the nature of the dynamical phases. We foresee many other applications of tensor networks in classical stochastic dynamics, including when the dynamical transition is continuous rather than first-order, and in the study of systems in dimension larger than one. More broadly, the crossover of ideas and techniques between quantum many-body and classical stochastics remains a fruitful area of investigation.

\bigskip

\noindent{\bf \em Acknowledgements --} 
This work was supported by 
the Deutsche Forschungsgemeinschaft (DFG, German Research Foundation) under Germany's Excellence Strategy -- EXC-2111 -- 390814868, by EPSRC Grant No.\ EP/R04421X/1 and by the Leverhulme Trust Grant No.\ RPG-2018-181. We acknowledge the hospitality of the Kavli Institute for Theoretical Physics at the University of California, Santa Barbara, where this work was started, and support from the National Science Foundation under Grant No. NSF PHY-1748958.

\bibliographystyle{apsrev4-1}
\bibliography{MPS-East}

\newpage
\onecolumngrid
\section{Supplemental Material} 

\renewcommand{\thefigure}{S\arabic{figure}}
\renewcommand{\theequation}{S\arabic{equation}}
\setcounter{equation}{0}
\setcounter{figure}{0}

\section{Numerical method}
\label{app:num}

The main MPS algorithm employed for this work is the variational optimization of a MPS with open boundary conditions, in order to solve
the minimization
\begin{equation}
\ket{\Psi}=\mathrm{argmin} \frac{\bra{\Psi}H\ket{\Psi}}{\bra{\Psi}\Psi\rangle},
\label{eq:varprin}
\end{equation}
over the set of MPS with fixed bond dimension $D$. The solution is the MPS
approximation to the ground state of the Hamiltonian $H$.
There are many reviews in the literature describing the development, technical details and applications of tensor network algorithms
like this one, as well as their extensions
to infinite systems, finite temperature and dynamics, and possible extensions to higher dimensions
~\cite{Schollwoeck2011,Verstraete2008}. 

It is convenient to express the algorithm fully in terms of tensor networks, by writing the Hamiltonian as a 
matrix product operator (MPO)~\cite{McCulloch2007,pirvu10mpo}, i.e. a MPS vector in the tensor product basis of  
operators (that is, as a linear combination of products of Pauli matrices).
Local Hamiltonians as the ones considered in this work have an exact MPO expression with small constant
bond dimension $D_H$ that does not depend on the system size.
Evaluating its expectation value in a MPS of bond dimension $D$, which is the fundamental ingredient for the variational minimization of the energy, 
has then a cost that scales as $O(d D_H D^3)$ in terms of the tensor dimensions, and linearly with the system size. This is crucial for the efficiency of the variational algorithm.
In particular, we can write the East Hamiltonian model for open boundary conditions with bond dimension $D_H=3$  (or 4 for periodic chains)
and the FA Hamiltonian with $D_H=4$ (or 6 for periodic boundary conditions).

\begin{figure}[h]
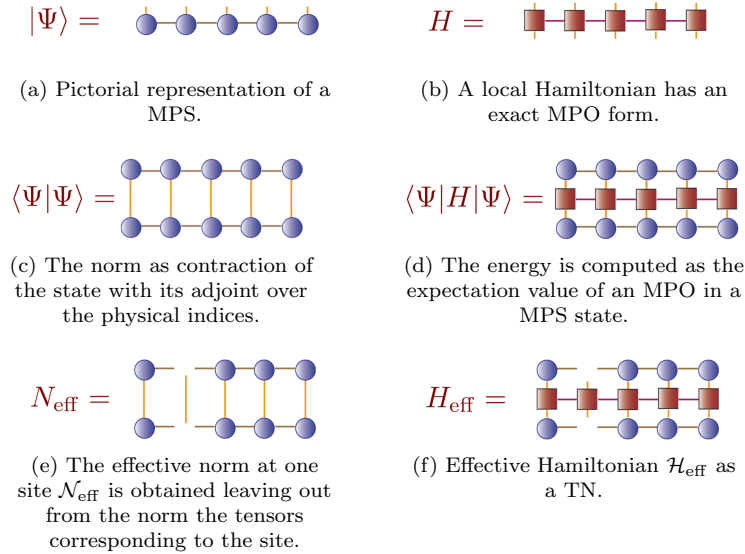

\centering
\centering
\subfloat[Pictorial representation of a MPS.]{\label{fig:MPS}\includegraphics[width=.24\columnwidth]{figuresSuppl/{{sketchTN_mps}}}}
\hspace{.05\columnwidth}
\subfloat[A local Hamiltonian has an exact MPO form.]{\label{fig:MPO}\includegraphics[width=.24\columnwidth]{figuresSuppl/{{sketchTN_mpo}}}}\\
\subfloat[The norm as contraction of the state with its adjoint over the physical indices.]{\label{fig:MPS}\includegraphics[width=.24\columnwidth]{figuresSuppl/{{sketchTN_norm}}}}
\hspace{.05\columnwidth}
\subfloat[The energy is computed as the expectation value of an MPO in a MPS state.]{\label{fig:MPO}\includegraphics[width=.26\columnwidth]{figuresSuppl/{{sketchTN_energy}}}}\\
\subfloat[The effective norm at one site $\mathcal{N}_{\mathrm{eff}}$ is obtained leaving out from the norm the tensors corresponding to the site.]{\label{fig:Neff}\includegraphics[width=.24\columnwidth]{figuresSuppl/{{sketchTN_neff}}}}
\hspace{.05\columnwidth}
\subfloat[Effective Hamiltonian $\mathcal{H}_{\mathrm{eff}}$ as a TN.]{\label{fig:Heff}\includegraphics[width=.24\columnwidth]{figuresSuppl/{{sketchTN_heff}}}}\\
\caption{Tensor networks and their contractions can be represented in a convenient pictorial language,
which simplifies the description of algorithms and operations. 
A solid geometrical form (e.g. circles or squares above) represents a tensor, 
with as many indices as depicted legs.
A contraction of two tensors over a certain index is represented as a connecting line.
The pictures show the graphical representation of MPS, MPO and their contractions, as they appear in the variational algorithm. }
\label{fig:TNsketch}
\end{figure}

The variational optimization then proceeds by fixing all tensors of the ansatz
\begin{equation}
|\Psi\rangle=\sum_{i_1,\ldots i_N=1}^d \mathrm{tr}\left ( A_1^{i_1} A_2^{i_2}\ldots A_N^{i_N} \right ) |i_1 i_2 \ldots i_N\rangle 
\end{equation}
but the one for site $k$, $A_k$, and 
rewriting the optimization~\eqref{eq:varprin}
as a local problem in terms of the single variable tensor. 
The local problem boils down to a generalized eigenvalue problem for the vectorized tensor,
$\mathcal{H}_{\mathrm{eff}} A_k=\lambda_{\min}\mathcal{N}_{\mathrm{eff}}  A_k$
~\cite{Schollwoeck2011,Verstraete2008},
where $\mathcal{H}_{\mathrm{eff}}$ ($\mathcal{N}_{\mathrm{eff}}$) is an effective Hamiltonian (norm matrix) of dimension $d D^2 \times d D^2$,
obtained by contracting all tensors in $\bra{\Psi} H \ket{\Psi}$ and in $\bra{\Psi} \Psi\rangle$, except for $A_k$; see a pictorial representation in Fig.~\ref{fig:TNsketch}.
This problem can then be solved with a standard eigensolver from a linear algebra numerical package, and the minimum eigenvalue $\lambda_{\min}$
corresponds to the estimate of the ground state energy.
Using a sparse eigensolver allows to keep the cost scaling as $D^3$ and to deal with very large values of the bond dimension.
The $k$-th tensor is updated with the solution of the local optimization, and then the procedure is repeated for all the tensors in the chain, 
sweeping back and forth until a certain convergence criterion (typically on the energy) is met. 
The same algorithm can be used to find higher excited states by imposing that the solution is orthogonal to already found levels.
This can be imposed at the level of the local problem, without changing the scaling of the leading cost, which is 
always $\mathcal{O}(D^3)$ (and grows polynomially with the number of computed levels).

For a run with fixed bond dimension, the algorithm is guaranteed to converge, because it can only decrease the energy in every step, 
although it may do so to a local minimum.
To improve the precision, one increases the bond dimension of the ansatz, typically using the previous solution with smaller $D$ as initial guess.
In a typical application, the algorithm is repeatedly run with increasing bond dimension, until the energy of the state is converged
to the desired precision~\footnote{See e.g.~\cite{Schollwoeck2011} for more details on initialization, convergence criteria, etc.}.
A notorious case in which the algorithm is slow to converge is that of critical systems, where the ground state requires a
bond dimension that grows polynomially with the system size in order to achieve a fixed precision, a situation that is well understood
by DMRG practitioners.
But on the other hand, having a state that can be well approximated by a MPS does not guarantee convergence of the algorithm.
A large density of states also hinders convergence, as happens for instance when trying to approximate 
excited states in the middle of the spectrum.

\subsection{Convergence}

\begin{figure}[h]
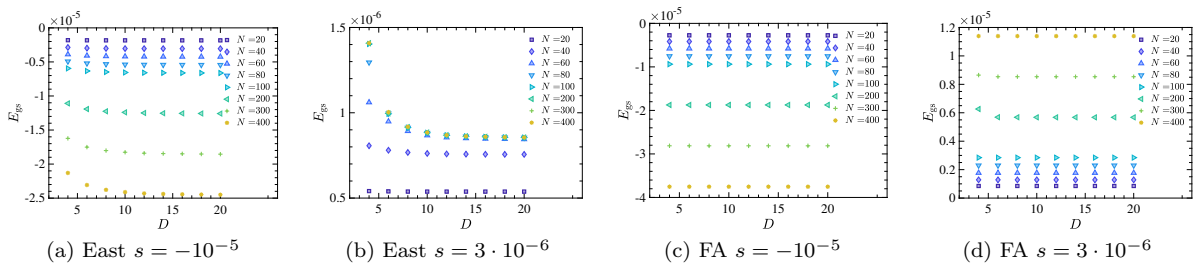

\centering
\subfloat[East $s=-10^{-5}$]{\label{fig:convEast_sneg_c0.05}\includegraphics[width=.2\columnwidth]{figuresSuppl/{{energyConvergence_East_c0.05_s-1e-05}}}}
\hspace{.02\columnwidth}
\subfloat[East $s=3\cdot 10^{-6}$]{\label{fig:convEast_spos_c0.05}\includegraphics[width=.2\columnwidth]{figuresSuppl/{{energyConvergence_East_c0.05_s3e-06}}}}
\hspace{.02\columnwidth}
\subfloat[FA $s=-10^{-5}$]{\label{fig:convFA_sneg_c0.05}\includegraphics[width=.2\columnwidth]{figuresSuppl/{{energyConvergence_FA_c0.05_s-1e-05}}}}
\hspace{.02\columnwidth}
\subfloat[FA $s=3\cdot 10^{-6}$]{\label{fig:convFA_spos_c0.05}\includegraphics[width=.2\columnwidth]{figuresSuppl/{{energyConvergence_FA_c0.05_s3e-06}}}}
\caption{Convergence of the energy as a function of the bond dimension in some of the most difficult cases ($c=0.05$ and small values of $s$).}
\label{fig:energyConv}
\end{figure}

\begin{figure}[h]
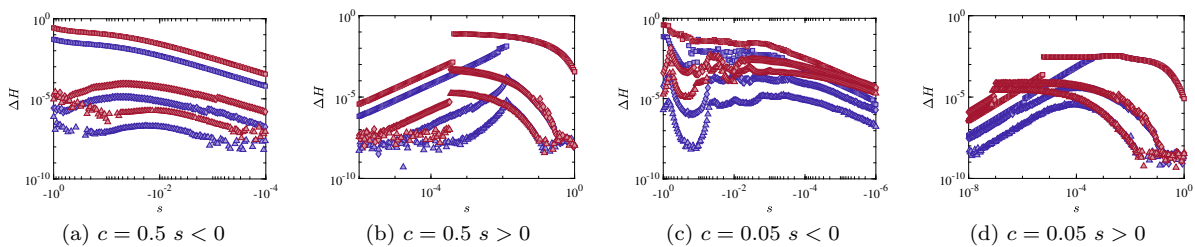

\centering
\subfloat[$c=0.5$ $s<0$]{\label{fig:Var_East_sneg_c0.5_D20}\includegraphics[width=.2\columnwidth]{figuresSuppl/{{plotVariance_East_c0.5_D20_sNeg}}}}
\hspace{.02\columnwidth}
\subfloat[$c=0.5$ $s>0$]{\label{fig:Var_East_spos_c0.5_D20}\includegraphics[width=.2\columnwidth]{figuresSuppl/{{plotVariance_East_c0.5_D20_sPos}}}}
\hspace{.02\columnwidth}
\subfloat[$c=0.05$ $s<0$]{\label{fig:Var_East_sneg_c0.05_D20}\includegraphics[width=.2\columnwidth]{figuresSuppl/{{plotVariance_East_c0.05_D20_sNeg}}}}
\hspace{.02\columnwidth}
\subfloat[$c=0.05$ $s>0$]{\label{fig:Var_East_spos_c0.05_D20}\includegraphics[width=.2\columnwidth]{figuresSuppl/{{plotVariance_East_c0.05_D20_sPos}}}}
\caption{Energy standard deviation (square root of variance)  $\Delta H=\sqrt{\langle H^2\rangle -\langle H\rangle^2}$ in the MPS approximation to the ground state for the East model. 
The plots show, for system sizes $N=20$ (blue) and $400$ (red) the systematic lowering of the variance as the bond dimension is varied from $D=2$ (squares), to $10$ (diamonds) and $20$ (triangles).
We show the detail of the most difficult regions, namely
the region of the plateaus for small $s<0$ (\ref{fig:Var_East_sneg_c0.5_D20} and \ref{fig:Var_East_sneg_c0.05_D20})
and the region of small $s>0$ around the phase transition (\ref{fig:Var_East_spos_c0.5_D20} and \ref{fig:Var_East_spos_c0.05_D20}).
By letting the algorithm run longer until a maximum bond dimension $D=100$, the variance of the peaks is reduced to $\Delta H \lesssim 10^{-5}$.}
\label{fig:varianceEast}
\end{figure}

\begin{figure}[h]
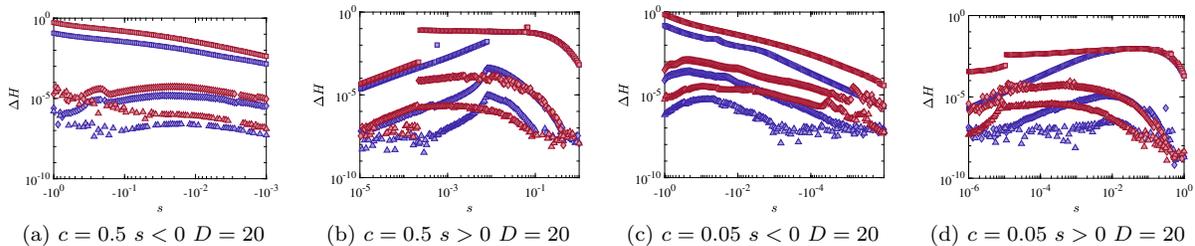

\centering
\subfloat[$c=0.5$ $s<0$ $D=20$]{\label{fig:Var_FA_sneg_c0.5_D20}\includegraphics[width=.2\columnwidth]{figuresSuppl/{{plotVariance_FA_c0.5_D20_sNeg}}}}
\hspace{.02\columnwidth}
\subfloat[$c=0.5$ $s>0$ $D=20$]{\label{fig:Var_FA_spos_c0.5_D20}\includegraphics[width=.2\columnwidth]{figuresSuppl/{{plotVariance_FA_c0.5_D20_sPos}}}}
\hspace{.02\columnwidth}
\subfloat[$c=0.05$ $s<0$ $D=20$]{\label{fig:Var_FA_sneg_c0.05_D20}\includegraphics[width=.2\columnwidth]{figuresSuppl/{{plotVariance_FA_c0.05_D20_sNeg}}}}
\hspace{.02\columnwidth}
\subfloat[$c=0.05$ $s>0$ $D=20$]{\label{fig:Var_FA_spos_c0.05_D20}\includegraphics[width=.2\columnwidth]{figuresSuppl/{{plotVariance_FA_c0.05_D20_sPos}}}}
\caption{Energy standard deviation (square root of variance) $\Delta H=\sqrt{\langle H^2\rangle -\langle H\rangle^2}$ in the MPS approximation to the ground state for the FA model. Qualitatively, we observe similar effects as for the East model, described in figure~\ref{fig:varianceEast}, 
As in  figure~\ref{fig:varianceEast}, 
we show  the systematic lowering of the variance as the bond dimension is varied from $D=2$ (squares), to $10$ (diamonds) and $20$ (triangles)
for system sizes $N=20$ (blue) and $400$ (red).
Again,
 bond dimension $D<100$ is enough to ensure very small variance over the most challenging range of parameters.}
\label{fig:varianceFA}
\end{figure}

We find the ground states over the largest part of the parameter space to be very well approximated by MPS with 
small bond dimension. 
The quality of the MPS approximation can be gauged from the convergence of observables as the bond dimension is increased.
We find this to be in general very fast, even for system sizes of several hundred sites.
We let the algorithm use bond dimensions as large as $D=100$, but in most of the cases analyzed, we find that a bond dimension $D=20$ is enough 
for the energy to be sufficiently converged.
As illustrated in figure \ref{fig:energyConv}, only in a few cases, mostly for small values of $c$ and around the phase transition, we find that a larger bond dimension allows us to reach a lower energy.
We also find that convergence becomes difficult for large systems when we
try to explore the region of the phase transition at small positive $s$ in both models. 
Since we have demonstrated that the states do not develop a large entropy, even in this region,
we attribute this behaviour to the density of states at the lowest energy becoming larger
for increasing system size.

A more accurate measure of how close the approximation is to an actual eigenstate is however provided by
the energy variance, $\Delta H^2=\langle H^2\rangle -\langle H\rangle^2$, which can be computed efficiently for any MPS.
In the cases studied in the paper, we find that a very small bond dimension, $D=20$, is already enough to obtain
a very small variance $\Delta H^2 \lesssim \mathcal{O}(10^{-10})$ for a wide range of values of $s$ and all system sizes up to $N=400$ (see the upper row of figures
\ref{fig:varianceEast} and \ref{fig:varianceFA}).
The exceptions are the region of the phase transition at small $s>0$ in both models, specially as $c$ decreases (see figures \ref{fig:Var_East_spos_c0.5_D20},
\ref{fig:Var_East_spos_c0.05_D20}, \ref{fig:Var_FA_spos_c0.5_D20} and \ref{fig:Var_FA_spos_c0.5_D20}).

\section{Detailed numerical results}
\label{app:extra}

\subsection{Finite size scaling of the active-inactive phase transition}
\label{app:extra1}

\begin{figure}[h]
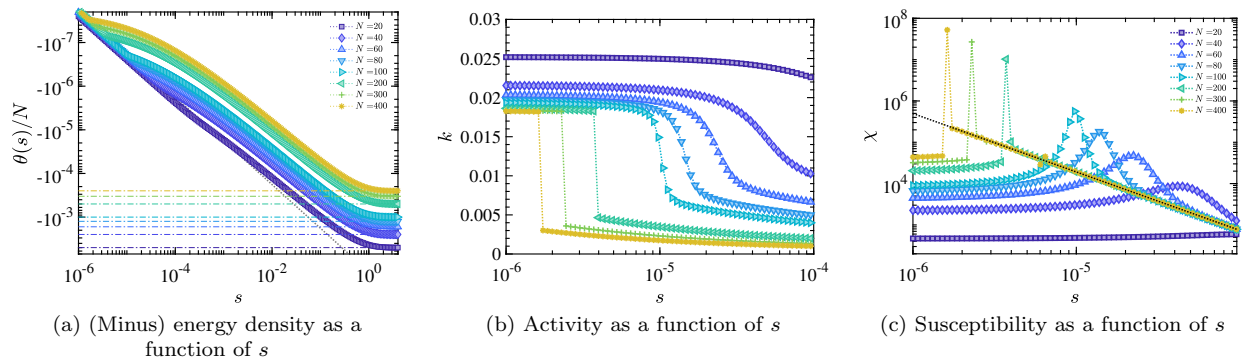

\centering
\subfloat[(Minus) energy density as a function of $s$]{\label{fig:Theta_East_c0.1}\includegraphics[width=.29\columnwidth]{figuresSuppl/{{plotTheta_East_c0.1}}}}
\hspace{.02\columnwidth}
\subfloat[Activity as a function of $s$]{\label{fig:K_East_c0.1}\includegraphics[width=.29\columnwidth]{figuresSuppl/{{plotActivity_East_c0.1}}}}
\hspace{.02\columnwidth}
\subfloat[Susceptibility as a function of $s$]{\label{fig:Chi_East_c0.1}\includegraphics[width=.28\columnwidth]{figuresSuppl/{{plotSusceptibility_East_c0.1}}}}
\caption{Scaling of the phase transition location for the East model with $c=0.1$.}
\label{fig:East_c0.1}
\end{figure}

\begin{figure}[h]
\centering
\subfloat[(Minus) energy density as a function of $s$]{\label{fig:Theta_FA_c0.1}\includegraphics[width=.29\columnwidth]{figuresSuppl/{{plotTheta_FA_c0.1}}}}
\hspace{.02\columnwidth}
\subfloat[Activity as a function of $s$]{\label{fig:K_FA_c0.1}\includegraphics[width=.29\columnwidth]{figuresSuppl/{{plotActivity_FA_c0.1}}}}
\hspace{.02\columnwidth}
\subfloat[Susceptibility as a function of $s$]{\label{fig:Chi_FA_c0.1}\includegraphics[width=.28\columnwidth]{figuresSuppl/{{plotSusceptibility_FA_c0.1}}}}
\caption{Scaling of the phase transition location for the FA model with $c=0.1$.}
\label{fig:FA_c0.1}
\end{figure}

\begin{figure}[h]
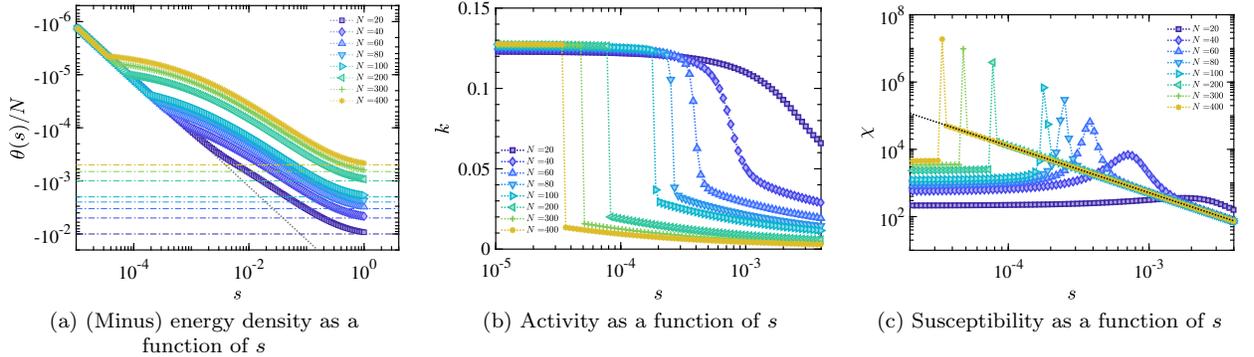

\centering
\subfloat[(Minus) energy density as a function of $s$]{\label{fig:Theta_FA_c0.2}\includegraphics[width=.29\columnwidth]{figuresSuppl/{{plotTheta_FA_c0.2}}}}
\hspace{.02\columnwidth}
\subfloat[Activity as a function of $s$]{\label{fig:K_FA_c0.2}\includegraphics[width=.29\columnwidth]{figuresSuppl/{{plotActivity_FA_c0.2}}}}
\hspace{.02\columnwidth}
\subfloat[Susceptibility as a function of $s$]{\label{fig:Chi_FA_c0.2}\includegraphics[width=.28\columnwidth]{figuresSuppl/{{plotSusceptibility_FA_c0.2}}}}
\caption{Scaling of the phase transition location for the FA model with $c=0.2$.}
\label{fig:FA_c0.2}
\end{figure}

To study the finite size scaling of the active-inactive phase transition we have simulated both models for varying system sizes $N\in[20,400]$
over a range of values for the equilibrium concentration $c\in [0.05,0.5]$.
We observe that all cases conform qualitatively to the behaviour discussed in the main text, and explicitly shown for the East model at $c=0.2$ [Fig. 1(a,b) in the main text].
The (normalized) SCGF, shown in Figs.~\ref{fig:Theta_East_c0.1}, \ref{fig:Theta_FA_c0.1} and \ref{fig:Theta_FA_c0.2},
which equals minus the energy density of the ground state, 
varies linearly with $s$, in agreement with the perturbative calculation around $s=0$. The grey dashed line in the figures shows the
linear response prediction in the thermodynamic limit, namely
$\theta/N =-2 c^2 (1-c) s$
for the East and
$\theta/N =-4 c^2 (1-c) s$
for the FA model.
The intersection of this line with the asymptotic value for $s\to\infty$, shown as dashed coloured lines in the plots, scales as $1/N$.
However, in the figures it is evident that the value of $s_c$ at which the actual crossing occurs can be orders of magnitude away from this prediction.

The activity 
\begin{equation}
k=-\frac{\theta'(s)}{N}=\frac{1}{N}\frac{d \Egs}{ds}
\end{equation}
can also be directly evaluated from the MPS ansatz, since
\begin{equation}
\frac{d E_{\mathrm{GS}}}{ds}=\bra{\Psi} \frac{dH}{ds}\ket{\Psi} ,
\end{equation}
only requires computing the expectation value
of local and two-body observables.
The results are shown in figures ~\ref{fig:K_East_c0.1}, \ref{fig:K_FA_c0.1} and \ref{fig:K_FA_c0.2}.
The numerical derivative of the activity yields the susceptibility $\chi=\frac{d\theta^2}{d s^2}$, shown in 
Figs.~\ref{fig:Chi_East_c0.1}, \ref{fig:Chi_FA_c0.1} and \ref{fig:Chi_FA_c0.2}.
The location $s_c$ of the phase transition for each system size is most precisely determined from
the position of the peak in $\chi$.

\bigskip 

\begin{figure}[h]
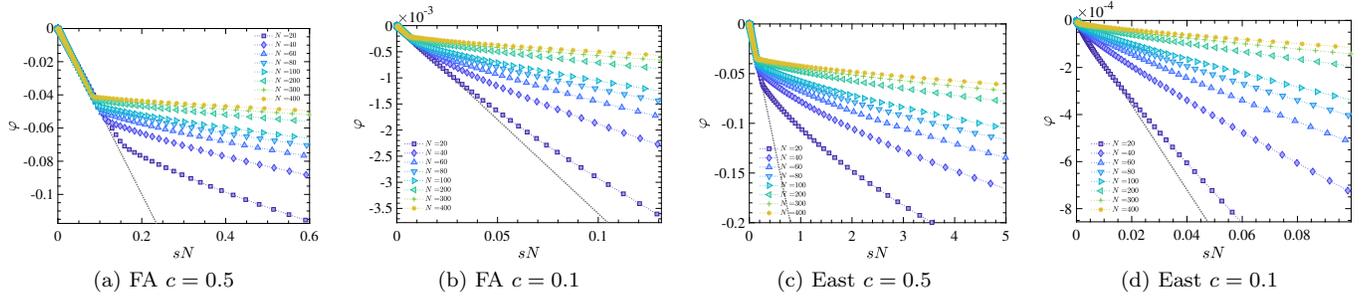

\centering
\subfloat[FA $c=0.5$]{\label{fig:theta-sN_FA_c0.5}\includegraphics[width=.23\columnwidth]{figuresSuppl/{{theta-sN_FA_c0.5}}}}
\hspace{.02\columnwidth}
\subfloat[FA $c=0.1$]{\label{fig:theta-sN_FA_c0.1}\includegraphics[width=.23\columnwidth]{figuresSuppl/{{theta-sN_FA_c0.1}}}}
\hspace{.02\columnwidth}
\subfloat[East $c=0.5$]{\label{fig:theta-sN_East_c0.5}\includegraphics[width=.23\columnwidth]{figuresSuppl/{{theta-sN_East_c0.5}}}}
\hspace{.02\columnwidth}
\subfloat[East $c=0.1$]{\label{fig:theta-sN_East_c0.1}\includegraphics[width=.23\columnwidth]{figuresSuppl/{{theta-sN_East_c0.1}}}}
\hspace{.02\columnwidth}
\caption{SCGF $\phi(\lambda)$ with $\lambda = s N$ in transition region, cf.\ Fig.~2 of Ref.\ \cite{Bodineau2012b}, for various sizes. The dashed line is the linear response behaviour.}
\label{fig:theta-sN}
\end{figure}

An alternative FSS analysis of the transition can be made by following the approach of Bodineau, Lecomte and Toninelli in Ref.~\cite{Bodineau2012b} (hereafter BLT) which considers in detail the FA model. BLT use the fact that within a region of size $1/N$ around the transition, that is for $s$ of order $1/N$ or equivalently for $\lambda = s N$ with $\lambda = O(1)$, the SCGF in terms of $\lambda$, $\phi(\lambda) = \theta(\lambda / N)$ is of $O(1)$ [rather than of $O(N)$ as for $s$ finite] and should progressively interpolate between two behaviours at large $N$ (see also \cite{Bodineau2012}). The two behaviours are that of  linear response, $\phi = - \langle k \rangle \lambda$, for $\lambda < \lambda_{c}$, and a   regime of constant $\phi = - \Sigma$ for $\lambda \geq \lambda_{c}$. Here $\Sigma$ is a ``surface tension'' related to the creation of an interface between the active and inactive phases, while $\langle k \rangle$ is the equilibrium activity per unit time. Figure \ref{fig:theta-sN} presents the SCGF in this representation for both models, cf.\ Fig.~2 of Ref.~\cite{Bodineau2012b}. The crossover between these two regimes is apparent both in the FA, as described by BLT, and also in the East model. Note that accessing the constant regime on the right is more difficult for lower $c$. 

\begin{figure}[h]
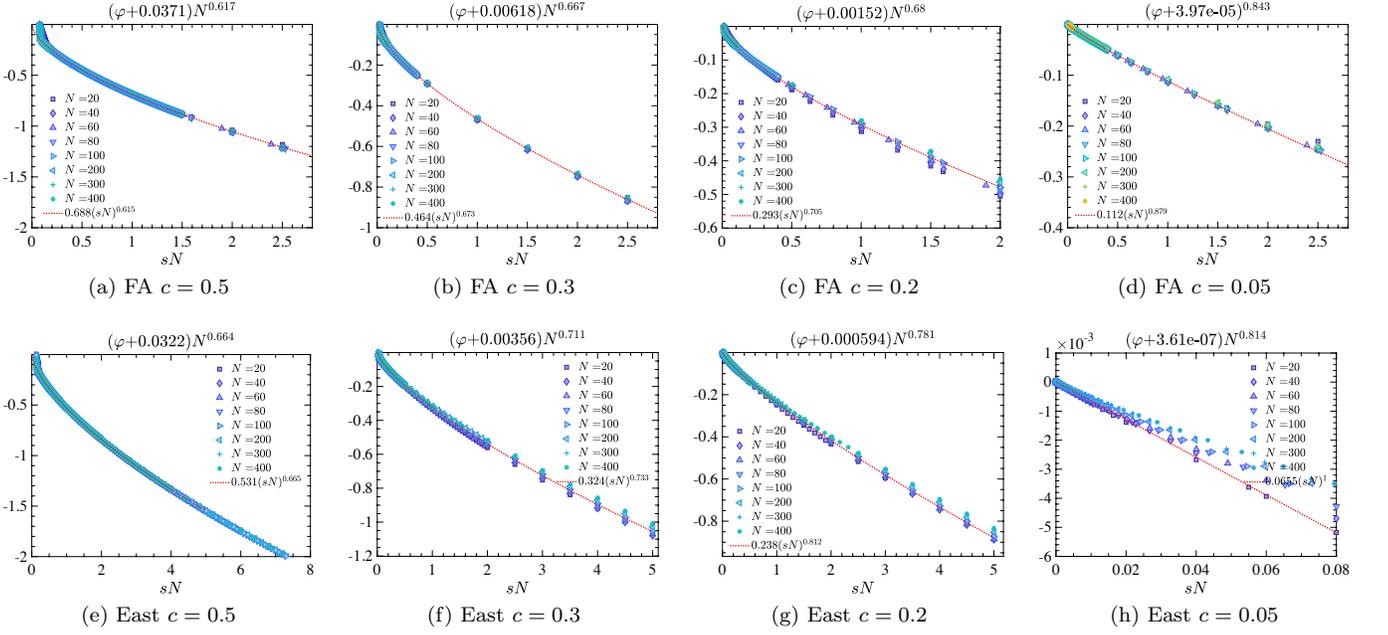

\centering
\subfloat[FA $c=0.5$]{\label{fig:collapse-theta-sN_FA_c0.5}\includegraphics[width=.23\columnwidth]{figuresSuppl/{{collapse-theta-sN_FA_c0.5}}}}
\hspace{.02\columnwidth}
\subfloat[FA $c=0.3$]{\label{fig:collapse-theta-sN_FA_c0.3}\includegraphics[width=.23\columnwidth]{figuresSuppl/{{collapse-theta-sN_FA_c0.3}}}}
\hspace{.02\columnwidth}
\subfloat[FA $c=0.2$]{\label{fig:collapse-theta-sN_FA_c0.2}\includegraphics[width=.23\columnwidth]{figuresSuppl/{{collapse-theta-sN_FA_c0.2}}}}
\hspace{.02\columnwidth}
\subfloat[FA $c=0.05$]{\label{fig:collapse-theta-sN_FA_c0.05}\includegraphics[width=.23\columnwidth]{figuresSuppl/{{collapse-theta-sN_FA_c0.05}}}}
\hspace{.02\columnwidth}
\subfloat[East $c=0.5$]{\label{fig:collapse-theta-sN_East_c0.5}\includegraphics[width=.23\columnwidth]{figuresSuppl/{{collapse-theta-sN_East_c0.5}}}}
\hspace{.02\columnwidth}
\subfloat[East $c=0.3$]{\label{fig:collapse-theta-sN_East_c0.3}\includegraphics[width=.23\columnwidth]{figuresSuppl/{{collapse-theta-sN_East_c0.3}}}}
\hspace{.02\columnwidth}
\subfloat[East $c=0.2$]{\label{fig:collapse-theta-sN_East_c0.2}\includegraphics[width=.23\columnwidth]{figuresSuppl/{{collapse-theta-sN_East_c0.2}}}}
\hspace{.02\columnwidth}
\subfloat[East $c=0.05$]{\label{fig:collapse-theta-sN_East_c0.05}\includegraphics[width=.23\columnwidth]{figuresSuppl/{{collapse-theta-sN_East_c0.05}}}}
\hspace{.02\columnwidth}
\caption{Collapse of SCGF $\phi(\lambda)$ for $\lambda > \lambda_{c}$, cf.\ Fig.~3 of Ref.\ \cite{Bodineau2012b}. }
\label{fig:collapse-theta-sN}
\end{figure}

\begin{figure}[h]
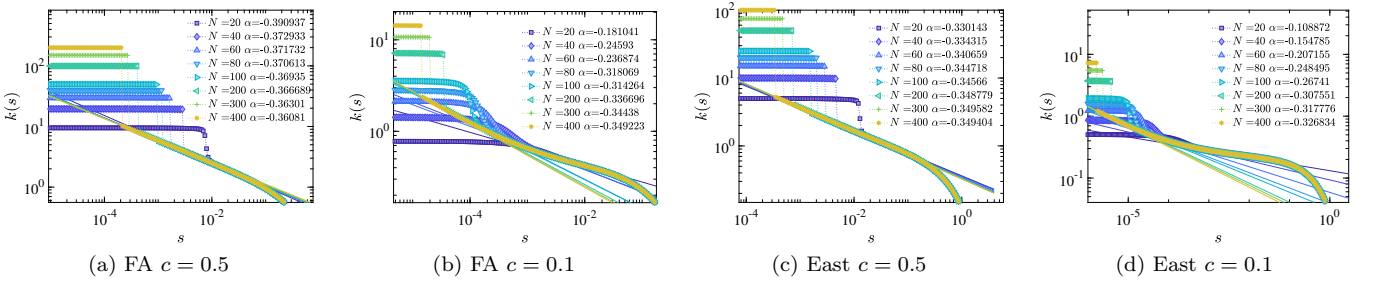

\centering
\subfloat[FA $c=0.5$]{\label{fig:collapseActivity_FA_c0.5}\includegraphics[width=.23\columnwidth]{figuresSuppl/{{collapseActivity_FA_c0.5}}}}
\hspace{.02\columnwidth}
\subfloat[FA $c=0.1$]{\label{fig:collapseActivity_FA_c0.1}\includegraphics[width=.23\columnwidth]{figuresSuppl/{{collapseActivity_FA_c0.1}}}}
\hspace{.02\columnwidth}
\subfloat[East $c=0.5$]{\label{fig:collapseActivity_East_c0.5}\includegraphics[width=.23\columnwidth]{figuresSuppl/{{collapseActivity_East_c0.5}}}}
\hspace{.02\columnwidth}
\subfloat[East $c=0.1$]{\label{fig:collapseActivity_East_c0.1}\includegraphics[width=.23\columnwidth]{figuresSuppl/{{collapseActivity_East_c0.1}}}}
\hspace{.02\columnwidth}
\caption{Activity $k(s)$. From the region where the curves collapse we can extract the exponent $\nu$ used in Fig.~\ref{fig:collapse-theta-sN}. We only show two values of $c$ for comparison, but this procedure was used to extract $\nu$ for all other $c$.}
\label{fig:collapseActivity}
\end{figure}

The prediction of BLT is that $\phi(\lambda)+\Sigma$ should behave as $-A \lambda^{\nu}$ for $\lambda > \lambda_{c}$. Figure \ref{fig:collapse-theta-sN} shows such a scaling for both the FA model and the East model, cf.\ Fig.~3 of \cite{Bodineau2012b}. We have estimated the exponent $\nu$ and the constant $\Sigma$ in the following way. Since the activity is (minus) the first derivative of the SCGF, in the relevant region it should scale as $\lambda^{\nu-1}$ [cf.\ the scaling of the susceptibility in Fig.~1(b) of the main text]. This allows to obtain $\nu$ without the need to simultaneously fit $\Sigma$. As discussed above, the activity can be calculated efficiently as it corresponds to the contraction of an MPO with the MPS. Figure \ref{fig:collapseActivity} shows the activity for both models. The smaller $c$ is the larger the system size $N$ is required to be to accurately extract $\nu$. With the exponent $\nu$ in hand we can then estimate $\Sigma$ by subtracting the $\lambda$ dependence from $\phi$. This is illustrated in Fig.~\ref{fig:Sigmas}(a,b) for $c=0.5$ for both models. 

For the FA model, BLT found $\Sigma \approx 0.077$ for the surface tension at $c=0.5$. 
In our case, for the FA model we find $\Sigma \approx 0.077/2$ at $c=0.5$, see Fig.~\ref{fig:Sigmas}(a), the factor of a half coming from the fact that we use open boundary conditions (which allows a single interface to be created, in contrast to the periodic boundary condition case). This result seems to confirm the BLT prediction. For the East model we find a slightly lower value at $c=0.5$, $\Sigma \approx 0.032$, see Fig.~\ref{fig:Sigmas}(b). As Fig.~\ref{fig:Sigmas}(c) shows, we observe that $\Sigma$ decreases significantly with $c$, with $\Sigma$ seemingly going as $\Sigma \propto c^{\xi}$ with $\xi \approx 3.3$ for the FA model (and decreasing even faster with $c$ for the East model). For the exponent, BLT predicted $\nu = 2/3$. While for large $c$ this is compatible with our findings, see Fig.~\ref{fig:collapseActivity}, we seem to find that $\nu$ increases with decreasing $c$. Nevertheless, this discrepancy might be due to the fact that for smaller values of $c$ extracting both $\Sigma$ and $\nu$ gets progressively more difficult.

\begin{figure}[h]
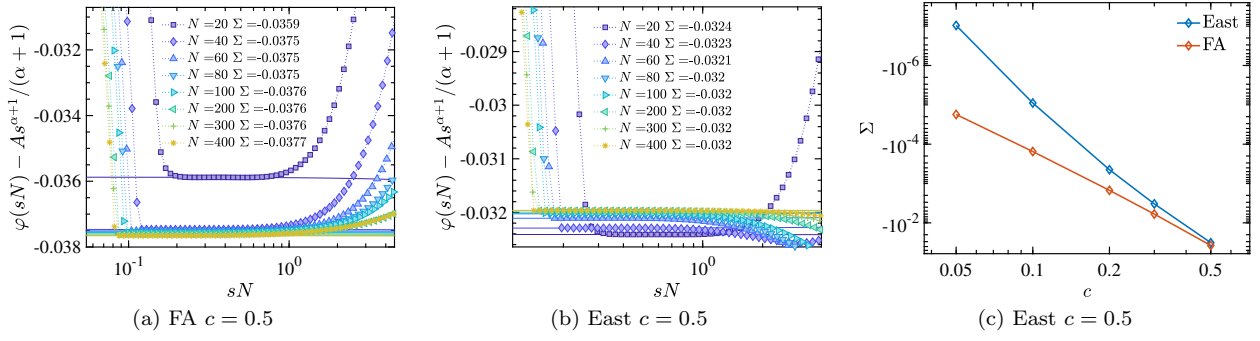

\centering
\subfloat[FA $c=0.5$]{\label{fig:SigmaFromcollapseActivity_FA_c0.5}\includegraphics[width=.29\columnwidth]{figuresSuppl/{{SigmaFromcollapseActivity_FA_c0.5}}}}
\hspace{.02\columnwidth}
\subfloat[East $c=0.5$]{\label{fig:SigmaFromcollapseActivity_East_c0.5}\includegraphics[width=.29\columnwidth]{figuresSuppl/{{SigmaFromcollapseActivity_East_c0.5}}}}
\hspace{.02\columnwidth}
\subfloat[East $c=0.5$]{\label{fig:Sigmas}\includegraphics[width=.29\columnwidth]{figuresSuppl/{{Sigmas}}}}
\hspace{.02\columnwidth}
\caption{(a) SCGF after subtracting the term proportional to $\lambda^{\nu}$ in order to estimate $\Sigma$ for the FA model. (b) Same for East model. (c) Surface tension $\Sigma$ as a function of $c$ for both models extracted by the procedure of panels (b,c) for all available $c$. For the FA model we get $\Sigma \propto c^{\xi}$ with $\xi \approx 3.3(2)$. }
\label{fig:Sigmas}
\end{figure}

\subsection{Structure of active phase}
\label{app:extra2}

\begin{figure}[h]
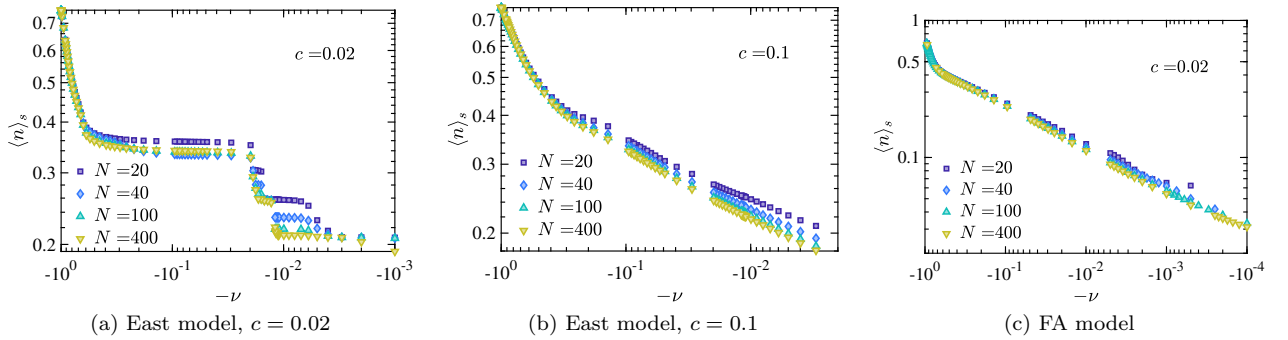

\centering
\subfloat[East model, $c=0.02$]{\label{fig:NvsNu_East_c0.02}\includegraphics[width=.3\columnwidth]{figuresSuppl/{{plotNvsNu_East_c0.02}}}}
\hspace{.02\columnwidth}
\subfloat[East model, $c=0.1$]{\label{fig:NvsNu_East_c0.1}\includegraphics[width=.29\columnwidth]{figuresSuppl/{{plotNvsNu_East_c0.1}}}}
\hspace{.02\columnwidth}
\subfloat[FA model]{\label{fig:NvsNu_FA_c0.02}\includegraphics[width=.29\columnwidth]{figuresSuppl/{{plotNvsNu_FA_c0.02}}}}
\caption{Average occupation in the active phase as a function of $-\nu=e^s-1$. For the East model (two leftmost plots), plateaus appear for small values of the equilibrium concentration, up to $c=0.1$ (central plot), where we only observe a cusp at $\nu\sim 0.14$ ($s\sim-0.13$). For larger values of $c$, the curve is smooth (see the inset of Fig. 2(a) in the main text), the same as for the FA model over the whole range of values of $c$.}
\label{fig:NvsNu_c0.1}
\end{figure}

\begin{figure}[h]
\centering
\subfloat[$c=0.02$]{\label{fig:dens_East_c0.02}\includegraphics[width=.23\columnwidth]{figuresSuppl/{{densitiesSurf_East_N20_c0.02}}}}
\hspace{.01\columnwidth}
\subfloat[$c=0.05$]{\label{fig:dens_East_c0.05}\includegraphics[width=.23\columnwidth]{figuresSuppl/{{densitiesSurf_East_N20_c0.05}}}}
\hspace{.01\columnwidth}
\subfloat[$c=0.1$]{\label{fig:dens_East_c0.1}\includegraphics[width=.23\columnwidth]{figuresSuppl/{{densitiesSurf_East_N20_c0.1}}}}
\hspace{.01\columnwidth}
\subfloat[$c=0.5$]{\label{fig:dens_East_c0.5}\includegraphics[width=.23\columnwidth]{figuresSuppl/{{densitiesSurf_East_N20_c0.5}}}}
\caption{Spatial distribution of density $n_i=(1-\langle \sigma_i^{z}\rangle)/2$ in the active phase of the East model  as a function of $s$ for a chain of length $N=20$ and increasing value of $c$.}
\label{fig:dens_East}
\end{figure}

\begin{figure}[h]
\centering
\subfloat[$c=0.02$]{\label{fig:dens_FA_c0.02}\includegraphics[width=.23\columnwidth]{figuresSuppl/{{densitiesSurf_FA_N20_c0.02}}}}
\hspace{.01\columnwidth}
\subfloat[$c=0.05$]{\label{fig:dens_FA_c0.05}\includegraphics[width=.23\columnwidth]{figuresSuppl/{{densitiesSurf_FA_N20_c0.05}}}}
\hspace{.01\columnwidth}
\subfloat[$c=0.1$]{\label{fig:dens_FA_c0.1}\includegraphics[width=.23\columnwidth]{figuresSuppl/{{densitiesSurf_FA_N20_c0.1}}}}
\hspace{.01\columnwidth}
\subfloat[$c=0.5$]{\label{fig:dens_FA_c0.5}\includegraphics[width=.23\columnwidth]{figuresSuppl/{{densitiesSurf_FA_N20_c0.5}}}}
\caption{Spatial distribution of density $n_i=(1-\langle \sigma_i^{z}\rangle)/2$ in the active phase of the FA model  as a function of $s$ for a chain of length $N=20$ and increasing value of $c$.}
\label{fig:dens_FA}
\end{figure}

\begin{figure}[h]
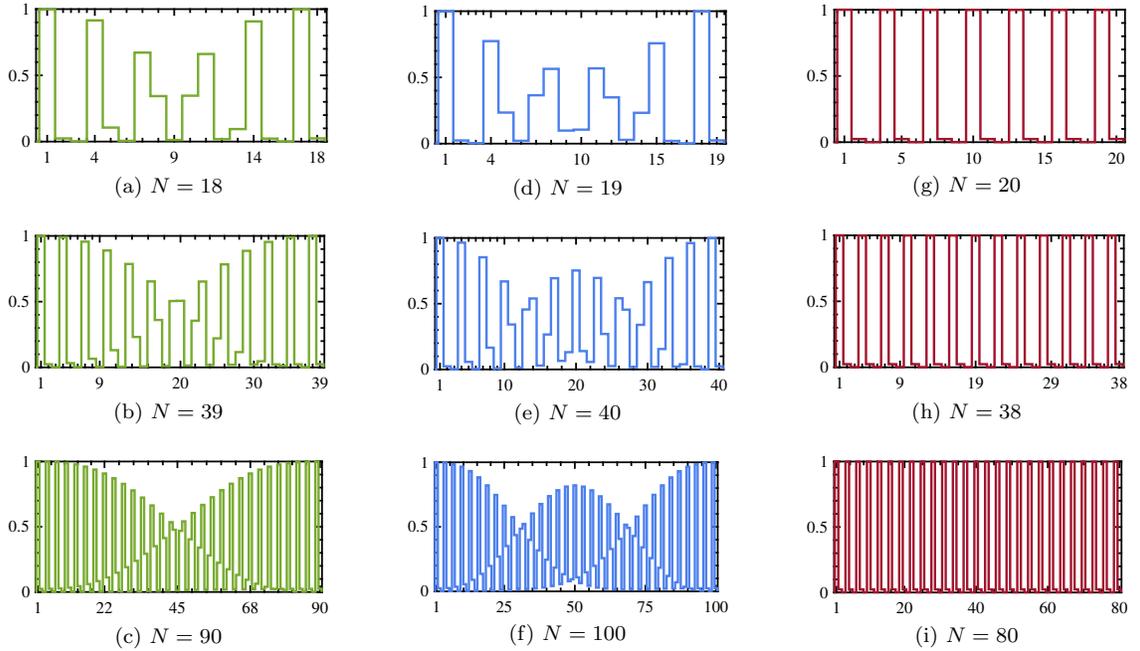

\centering
\begin{minipage}{.28\columnwidth}
\subfloat[$N=18$]{\label{fig:densDistr_East_mod0}\includegraphics[width=.85\columnwidth]{figuresSuppl/{{occupationVsX_East_N18_nu0.095_c0.02}}}}\\
\subfloat[$N=39$]{\label{fig:densDistr_East_mod0}\includegraphics[width=.85\columnwidth]{figuresSuppl/{{occupationVsX_East_N39_nu0.095_c0.02}}}}\\
\subfloat[$N=90$]{\label{fig:densDistr_East_mod0}\includegraphics[width=.85\columnwidth]{figuresSuppl/{{occupationVsX_East_N90_nu0.095_c0.02}}}}
\end{minipage}%
\hspace{.01\columnwidth}
\begin{minipage}{.28\columnwidth}
\subfloat[$N=19$]{\label{fig:densDistr_East_mod1}\includegraphics[width=.85\columnwidth]{figuresSuppl/{{occupationVsX_East_N19_nu0.095_c0.02}}}}\\
\subfloat[$N=40$]{\label{fig:densDistr_East_mod1}\includegraphics[width=.85\columnwidth]{figuresSuppl/{{occupationVsX_East_N40_nu0.095_c0.02}}}}\\
\subfloat[$N=100$]{\label{fig:densDistr_East_mod1}\includegraphics[width=.85\columnwidth]{figuresSuppl/{{occupationVsX_East_N100_nu0.095_c0.02}}}}\\
\end{minipage}%
\hspace{.01\columnwidth}
\begin{minipage}{.28\columnwidth}
\subfloat[$N=20$]{\label{fig:densDistr_East_mod2}\includegraphics[width=.85\columnwidth]{figuresSuppl/{{occupationVsX_East_N20_nu0.095_c0.02}}}}\\
\subfloat[$N=38$]{\label{fig:densDistr_East_mod2}\includegraphics[width=.85\columnwidth]{figuresSuppl/{{occupationVsX_East_N38_nu0.095_c0.02}}}}\\
\subfloat[$N=80$]{\label{fig:densDistr_East_mod2}\includegraphics[width=.85\columnwidth]{figuresSuppl/{{occupationVsX_East_N80_nu0.095_c0.02}}}}
\end{minipage}
\caption{Distribution of density along the chain for $c=0.02$ and $s=-0.1$ (in the middle of the $\langle n\rangle_s=1/3$ plateau in Fig.~\ref{fig:dens_East_c0.02})
depending on the congruence of the chain length modulo $3$ being $0$ (left column), $1$ (center) or $2$ (right column).}
\label{fig:densDistrib_East}
\end{figure}

As discussed in the main text, the active phase of both models exhibits very different features, which we can explore with our results. 
In the East model, for small values of $c$, we recover the hierarchy of plateaus with well defined average density
predicted in~\cite{Jack2013},  extending between values of $\nu=1-e^s$ equal to integer powers of $c$.
This is clearly appreciated in Fig.~\ref{fig:NvsNu_East_c0.02} for $c=0.02$, where the plateau at $\langle n\rangle=1/3$ which ends at $\nu=0.02$
is already converged in system size.
For $c=0.1$ there is no plateau structure anymore, but a cusp remains in the average density at $\nu\sim 0.14$, as shown in Fig.~\ref{fig:NvsNu_East_c0.1}, 
while for yet larger values of $c$, the curve is smooth (see for instance the inset of figure 2(a) in the main text).
For the FA model, on the other hand, there are no similar features in the average density, as shown explicitly by 
Fig.~\ref{fig:NvsNu_FA_c0.02} and figure 2(b) in the main text.

To explore in more detail the structure of the active phase in both models, we have computed the spatial dependence of the density across the same 
range of values of $s$ spanned by Fig.~\ref{fig:NvsNu_c0.1}.
We show the results for a system size $N=20$ in figures~\ref{fig:dens_East} (for the East) and ~\ref{fig:dens_FA} (for the FA model).
In the case of the East model, the figure shows how the fixed average density of the plateaus is achieved by means of a
regular modulation of the local density. For the $\langle n\rangle_s=1/3$ plateau, the state has one occupied site followed by two almost empty ones,
a structure that the density plots in~\ref{fig:dens_East} clearly show.
The extension of the plateau decreases, and the position of its boundary moves to larger $s$ as $c$ increases, and 
they have disappeared completely at $c=0.5$ (fig.~\ref{fig:dens_East_c0.5}).
For the FA model, on the other hand, no plateaus occur for any value of $c$, as explicitly shown by Fig.~\ref{fig:dens_FA} for $c\in[0.02,0.5]$.

The period three modulation of density shown in Fig.~\ref{fig:dens_East} for the case of $N=20$ is present also in larger systems, 
but the detailed structure depends on the congruence modulo three of the system size.
This is shown explicitly in Fig.~\ref{fig:densDistrib_East}, for the particular case $c=0.02$, $s=-0.1$ and different system sizes.

\subsection{Entanglement}
\label{app:extra3}

\begin{figure}[h]
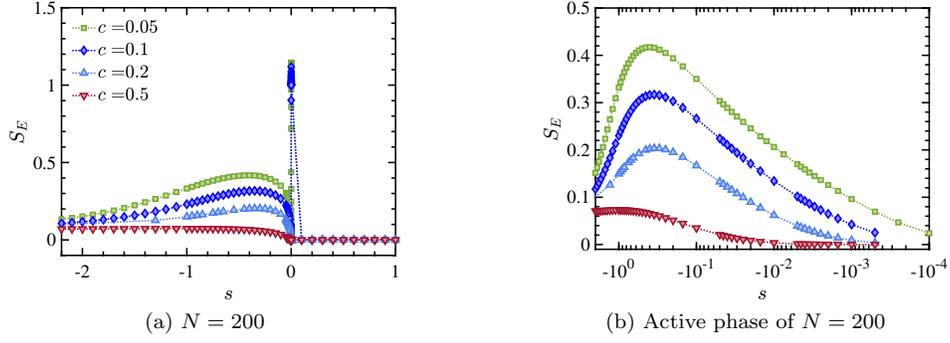

\centering
\subfloat[$N=200$]{\label{fig:entropy_FA_N200}\includegraphics[width=.29\columnwidth]{figuresSuppl/{{plotEntropy_FA_N200}}}}
\hspace{.1\columnwidth}
\subfloat[Active phase of $N=200$]{\label{fig:entropy_FA_N200_active}\includegraphics[width=.3\columnwidth]{figuresSuppl/{{plotEntropy_FA_N200_active}}}}
\caption{Entropy in the ground state of the FA model. The left plot shows the overall behaviour with respect to $s$ for a chain of size $N=200$ and different values of $c$, and the right one shows the detail of the active phase.}
\label{fig:entropy_FA}
\end{figure}

\begin{figure}[h]
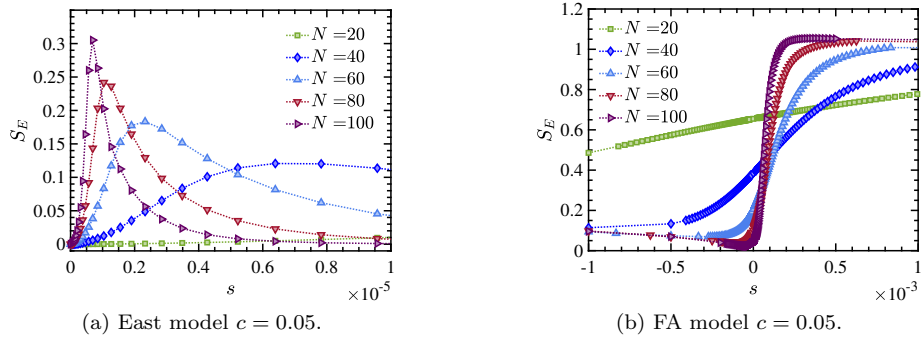

\centering
\subfloat[East model $c=0.05$.]{\label{fig:entropys0_East}\includegraphics[width=.285\columnwidth]{figuresSuppl/{{plotEntropyAt0_East_c0.05}}}}
\hspace{.1\columnwidth}
\subfloat[FA model $c=0.05$.]{\label{fig:entropys0_FA}\includegraphics[width=.285\columnwidth]{figuresSuppl/{{plotEntropyAt0_FA_c0.05}}}}
\caption{Entropy around $s=0$ in the ground state of the East and FA model at $c=0.05$ for different system sizes. }
\label{fig:entropy_s0}
\end{figure}

\begin{figure}[h]
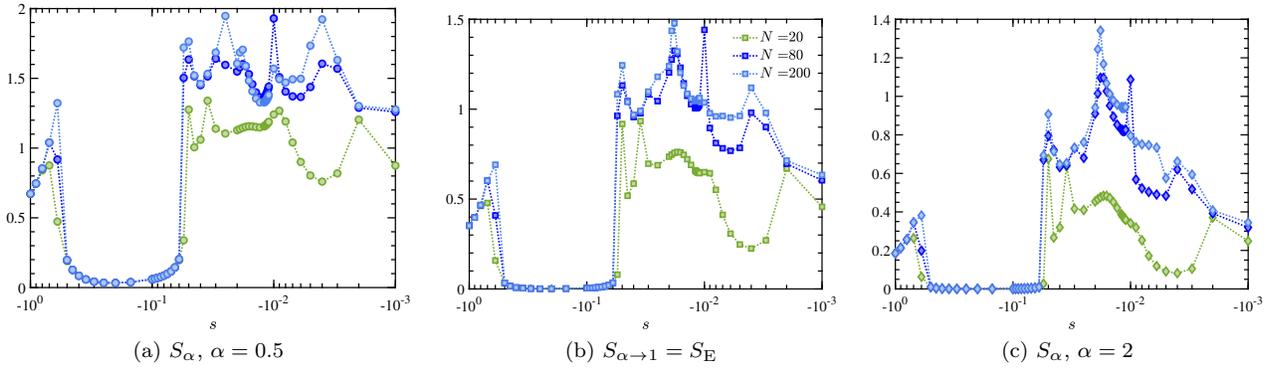

\centering
\subfloat[$S_{\alpha}$, $\alpha=0.5$]{\label{fig:renyiEast_0.5}\includegraphics[width=.3\columnwidth]{figuresSuppl/{{plotRenyi_East_alpha0.5_c0.05}}}}
\hspace{.02\columnwidth}
\subfloat[$S_{\alpha\to 1}=\SE$ ]{\label{fig:renyiEast_1}\includegraphics[width=.29\columnwidth]{figuresSuppl/{{plotRenyi_East_alpha1_c0.05}}}}
\hspace{.02\columnwidth}
\subfloat[$S_{\alpha}$, $\alpha=2$]{\label{fig:renyiEast_2}\includegraphics[width=.29\columnwidth]{figuresSuppl/{{plotRenyi_East_alpha2_c0.05}}}}
\caption{Entanglement in the active phase of the East model, as measured by the von Neumann and Renyi entropies $S_{\alpha}$ of the half-chain
for $c=0.05$ and various system sizes (congruent to 2 modulo 3).}
\label{fig:RenyiEast_c0.05}
\end{figure}

The states we are looking for contain very little entanglement.
We have shown explicitly in the main text that the entanglement entropy with respect to a division of the chain in two $\SE$
remains bounded by a constant even in the region of the phase transition for the East model [see Fig. 3 in the main text].
The same is true in the case of the FA model, as shown in Fig.~\ref{fig:entropy_FA_N200} for a chain of length $N=200$.
As one can expect from the previous discussion, in this case, no peaks of the entropy occur within the active phase,
and the entropy is smooth for all values of $c$; see zoom in Fig.~\ref{fig:entropy_FA_N200_active}, to be compared to Fig. 3(b) in the main text.
The maximum of the entropy occurs, instead, around the phase transition, for small positive values of $s$,
but the magnitude of the peak shows only a mild dependence on the system size, similar to what we observed in the East model.

There are however differences between both models in the behaviour of the entropy at very small $s$, as shown in Fig.~\ref{fig:entropy_s0}. 
In the East model, the entropy at $s=0$ is strictly zero, corresponding to a product ground state [see Eq.~(4) of the main text and Fig.~\ref{fig:entropys0_East}], and 
builds up to a peak around the transition. In the case of the FA model, the ground state at $s=0$, in the subspace orthogonal to the state with
no excitations, is not a product state [see Eq.~(3) of the main text], and thus the entropy is not exactly zero, but has a finite value at $s=0$, namely
\begin{equation}
\SE^{\mathrm{FA}}(s=0)=\mathrm{H}\left( \frac{1}{2}
\left [ 1+\sqrt{1-\left(\frac{2}{1+(1-c)^{-N/2}}\right)^2}\right]\right),
\end{equation}
where $\mathrm{H}(x)=-x \log_2 x -(1-x) \log_2(1-x)$ is the Shannon entropy.
The value of $\SE^{\mathrm{FA}}(s=0)$ decreases fast with the system size $N$ (Fig.~\ref{fig:entropys0_FA}). For $s>0$, $\SE$ grows also in the case of the FA model, towards a value of order one. 
Afterwards we find an almost vanishing energy.
Notice that to the right of the transition, in the limit $s\to \infty$, the ground state is doubly degenerate, 
and the component that is symmetric under parity, thus in the same subspace as the ground state at $s=0$, would have entanglement $\SE=1$.
However, at sufficiently large $s$, the algorithm prefers to break the symmetry to find a solution with smaller bond dimension.

The bounded entropy alone is not enough to guarantee the approximability of the ground state by a MPS~\cite{Schuch2008}.
To gather more compelling evidence we can also study the scaling of Renyi entropies, 
defined as  $S_\alpha=\mathrm{Tr}\rho^{\alpha}/(1-\alpha)$ for $\alpha>0$, and which in the limit
$\alpha \to 1$ converge to the von Neumann entropy.
 An area law for a $S_{\alpha}$ with $\alpha<1$ would imply approximability of the state as a MPS, as demonstrated in \cite{Schuch2008}.
The MPS ansatz gives natural access to the Schmidt values for any cut of the chain,
so that  all $S_{\alpha}$ can be computed efficiently.
We show the values of the von Neumann and Renyi entropies for the active phase of the East model with $c=0.05$ 
in the region of the plateaus (since it is the region with the largest entropy we found), in Fig.~\ref{fig:RenyiEast_c0.05}. 
Again we see that, although the magnitude of the peaks is not fully converged, the dependence on the system size
is very mild.

\end{document}